\documentclass[11pt,twoside]{article}
\usepackage{asp2010}

\resetcounters

\def \othreec {[O{\sc iii}] }

\def \ha {H$\alpha$}

\def \vhel{\ifmmode{~V_{{\rm HEL}}}\else{~$V_{{\rm HEL}}$}\fi}
\def\kms{\ifmmode{~{\rm km\,s}^{-1}}\else{~km s$^{-1}$}\fi}

\def \arcdeg{$^\circ$\ }

\begin{document}

\title{A new [O\textsc{iii}] $\lambda$5007~\AA\ Galactic Bulge Planetary Nebula Luminosity Function}
\author{A. V. Kovacevic$^1$, Q. A. Parker$^{1,2}$, G. H. Jacoby$^3$ and B. Miszalski$^4$
\affil{$^1$Macquarie University, Sydney, Australia}
\affil{$^2$Anglo-Australian Observatory, Sydney, Australia}
\affil{$^3$GMTO Corporation, CA, USA}
\affil{$^4$University of Hertfordshire, Hatfield, UK}}

\begin{abstract}
The Planetary Nebulae Luminosity Function (PNLF) describes the collective luminosity evolution for a given population of Planetary Nebulae (PN). A major paradox in current PNLF studies is in the universality of the absolute magnitude of the brightest PNe with galaxy type and age. The progenitor central-star mass required to produce such bright PNe should have evolved beyond the PNe phase in old, red elliptical galaxies whose stellar populations are $\sim$10~Gyr. Only by dissecting this resolved population in detail can we attempt to address this conundrum. The Bulge of our Galaxy is predominantly old \citep{Z03} and can therefore be used as a proxy for an elliptical galaxy, but with the 
significant advantage that the population is resolvable from ground based telescopes. We have used the MOSAIC-II camera on the Blanco 4-m at CTIO to carefully target $\sim$80~square degrees of the Galactic Bulge and establish accurate \othreec fluxes for 80\% of Bulge PNe currently known from the Acker and MASH catalogues. Construction of the \othreec Bulge PNLF has allowed us to investigate placement of PNe population sub-sets according to morphology and spectroscopic properties the PNLF and most importantly, whether any population subset might constitute the bright end of the LF. Our excellent, deep data also offers exciting prospects for significant new PNe discoveries and \othreec morphological studies. 
\end{abstract}

\section{Introduction}
\label{sec:intro}

For the past $\sim$20~yrs, the \othreec PNLF 
has developed into a well-established and widely used reliable extra-galactic distance indicator, permitting 
accurate distance measurement for galaxies out to the Fornax and Virgo clusters \citep{JCF90,FJP07,G07}.  
\citet{C89} published the refined fit to the luminosity function:

\begin{equation}
N(M) \propto e^{0.307M}(1-e^{3(M^{*}-M)})
\label{eq:lf} \end{equation}

When the distance to the population is accounted for, the fit to the bright-end cut-off 
is found to have a precisely defined absolute 
magnitude of M$^{*}=-4.47^{+0.02}_{-0.03}$ \citep{C02}. Except for a weak 
metallicity dependence \citep{D92}, this bright-end cut-off is found to be invariant to galaxy type or age. 
This fundamental inconsistency was outlined by \citet{C05}. 

The central stars of the brightest planetaries have luminosities of $\sim$6000~L$_{\sun}$, which 
infers masses of $>$0.6~M$_{\sun}$ \citep{2004A&A...423..995M} and 
consequently that they evolved through the main-sequence with a mass $>$2~M$_{\sun}$ \citep{W00}. Stars with these 
relatively high masses only live for $\sim$1-2~Gyr and therefore we do not expect to detect such stars 
in old populations such as elliptical galaxies that are $\sim$10~Gyr old and have undergone no recent 
star formation. However, they are nonetheless observed in such old populations (for a more detailed discussion see 
\citealt{J97}).

This paradox has seriously inhibited our understanding and interpretation of the observed 
invariability of the PNLF bright-end cut-off to population age. 
We propose to address this problem via construction of a new, significantly deeper 
Bulge PNLF to 
identify whether the bright-end of the luminosity function 
is comprised of PN that have evolved from old, population-II stars via some peculiar path to 
enhanced luminosity, e.g. through binarity \citep{C05}, or whether 
it is primarily populated by younger, higher mass, bipolar nebulae of Type-I \citep{TPP97}.  

\section{Optical observations and correcting for extinction}
\label{sec:oiiiphotom}

We obtained \othreec photometry for 435 previously known and MASH PN \citep{P06,M08} in 
a 10\arcdeg$\times$10\arcdeg region toward the 
Galactic Bulge \citep{K10a}. 
From these data, we measured \othreec fluxes 
and angular diameters, both of which were in agreement to within the errors of the most 
trusted literature sources. 
Previously, only 48 PN in this region had published 
\othreec fluxes, so 
we are increasing the number of PN with fluxes by a factor of eight. 
These data provide 
accurate fluxes and diameters for the largest sample of PN 
towards this region of the Bulge ever published. 

Optical observations made toward the Galactic Bulge are subject to a large amount of 
patchy and highly variable interstellar absorption and scatter due to dust along the line-of-sight. 
It is therefore important to ascertain the magnitude of extinction towards PN individually. 
We note that by correcting for the amount of dust towards each PN, we are also correcting for the dust 
internal to the nebula shell. This is not done for extragalactic PNLF studies, where a singular global 
extinction correction is applied to all PN. Accounting for this in PN that have evolved from high-mass central stars, 
where their inherently large dust content would usually act to self-extinct their 
bright \othreec emission, may prove 
detrimental and place them at a significantly higher value than M$^{*}$. This has also been observed when correcting 
the extinction towards individual PN for PNLF construction in the Bulge of M31.

To deredden our fluxes we undertook 
spectroscopic observation of $\sim$400 PN towards the Galactic Bulge (see Miszalski et al., 2009a). 
For $\sim$350 of these objects which also had \othreec photometry, we 
obtained Balmer decrement measurements with which to derive extinction estimates. 
These data are given in \citet{K10b}.

\section{The new Galactic Bulge \othreec PNLF}

Previously defined criteria to identify and exclude disc PN from any Bulge PNLF construction 
are more assumption than fact, having never undergone rigorous testing. We re-evaluate and re-define the 
criteria required for PN Bulge membership in terms of location, radio flux, radial velocity, 
surface brightness and radio and optical angular diameters. 

\begin{figure}
\centering
\includegraphics[scale=0.7]{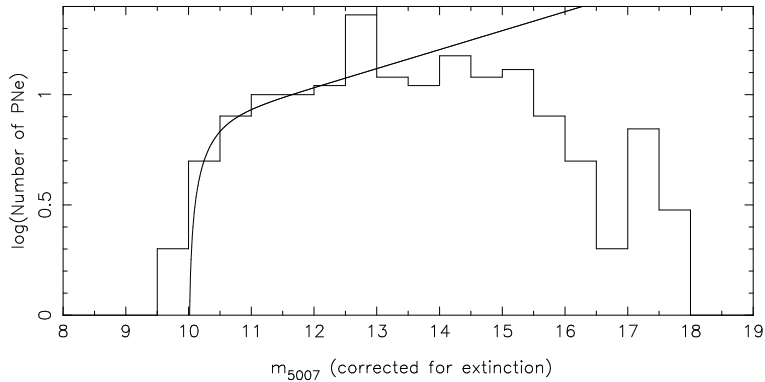}
\caption{The \othreec luminosity function for Galactic Bulge Planetary Nebulae. }
\label{fig:lf}
\end{figure}

With this disc-cleaned sample, we present a new, preliminary version of the \othreec PNLF (Fig.
\ref{fig:lf}). After fitting equation \ref{eq:lf} 
to the distribution, we calculate 
the bright-end cut-off has an absolute magnitide of $-4.38\pm0.13$, 
assuming a distance to the Galactic Centre of 7.52~kpc \citep{N06}. 
This agrees, to within the errors, with the canonical value of
M$^{*}=-4.47^{+0.02}_{-0.03}$ \citep{C02}.  
The high-resolution of the MOSAIC-II Imaging has permitted preliminary elucidation of morphology of some Bulge PN 
for the first time. This, together with SHS \ha\ images \citep{P05} 
and other imagery where it exists in the literature, 
will enable us to assign more conclusive morphological classifications from a multi-wavelength approach.  

Interestingly, within our sample, we find that 
neither Type I nor bipolar PN populate the bright-end in agreement with \citet{jacoby02}. 
Instead they dominate the faint end of the SMC PNLF, perhaps as 
a consequence of fast evolution and a high amount of dust present.

\section{Future work}

\begin{figure}
\centering
\includegraphics[scale=0.46]{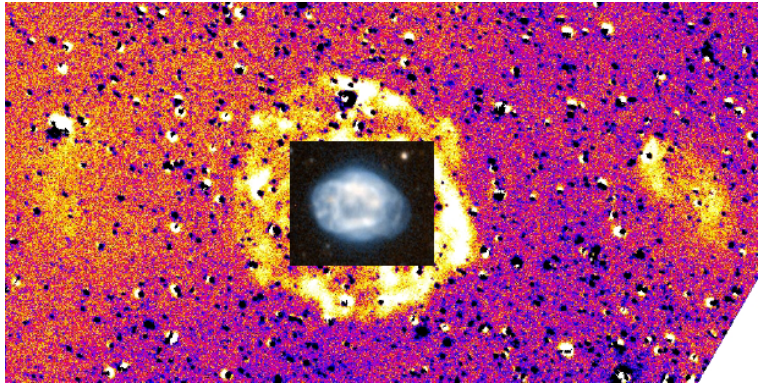}
\caption{A faint halo of emission and external arcs detected around IC 4673. 
Inset image taken from \citet{S92}. 
}
\label{fig:IC4673}
\end{figure}

Our \othreec data not only allows for accurate measurement of fluxes, diameters and 
identifcation of morphologies of PN, 
but also, the sensitivity attained has permitted the discovery of faint nebular extensions pertaining to 
previous periods of mass loss in some PN. This is exemplified by the case of IC 4673 
(see Fig. \ref{fig:IC4673}), where a halo surrounds the PN \citep{corradi03} together with 
external east-west arcs which could signifiy the presence of 
a binary central star \citep{M09b}.
The depth of the survey data has also allowed for discovery of numerous faint new PN candidates. 
The \othreec images of PN and extensions around PN will be made available online in the 
near future.

\section{Conclusions}

We provide accurate \othreec fluxes and angular diameters for the largest sample of PN 
in the 10\arcdeg$\times$10\arcdeg region toward the Galactic Bulge ever obtained. 
We obtained optical spectroscopy to provide accurate measurement 
and derivation of the reddening corrections towards $\sim$350 of the PN with \othreec fluxes.
The general criteria considered for Bulge membership was re-defined, and implemented to exclude any disc PN. 
This cleaned sample was then used to construct our preliminary new Galactic Bulge PNLF. 

This is the deepest Bulge PNLF ever constructed, ranging over 8 orders of magnitude 
and is complete to 4 magnitudes below M$^{*}$. 
The fit to the bright-end cut-off has an absolute magnitude of M$^{*} = -4.38\pm0.13$, 
using a distance of 7.52~kpc. After a brief analysis of high quality spectra 
from the literature, the PN constituents residing in the brightest bins 
are not found to be of Type-I or have bi-polar morphology. 

\acknowledgements 

AVK acknowledges Macquarie University for a PhD scholarship and 
the Astronomical Society of Australia for travel assistance.

\bibliography{references.bib}

\end{document}